\documentclass[sigconf]{acmart}

\usepackage[ruled,vlined, linesnumbered]{algorithm2e}
\usepackage{subfig}
\newcommand{\quotes}[1]{``#1''} 
\usepackage{tabularx}
\usepackage{graphicx}
\usepackage{adjustbox}
\usepackage{longtable}
\usepackage{color,soul}
\AtBeginDocument{%
  \providecommand\BibTeX{{%
    \normalfont B\kern-0.5em{\scshape i\kern-0.25em b}\kern-0.8em\TeX}}}

\setcopyright{acmcopyright}
\copyrightyear{2018}
\acmYear{2018}
\acmDOI{10.1145/1122445.1122456}

\acmConference[Woodstock '18]{Woodstock '18: ACM Symposium on Neural
  Gaze Detection}{June 03--05, 2018}{Woodstock, NY}
\acmBooktitle{Woodstock '18: ACM Symposium on Neural Gaze Detection,
  June 03--05, 2018, Woodstock, NY}
\acmPrice{15.00}
\acmISBN{978-1-4503-XXXX-X/18/06}

\begin{document}

\title{An Intelligent Recommendation-cum-Reminder System}




\author{Rohan Saxena}
\email{rohaan.saxena14@gmail.com}
\affiliation{%
 \institution{BIET Jhansi, India}
 \country{India}}

\author{Maheep}
\email{chaudhary.maheep28@gmail.com}
\affiliation{%
 \institution{BIET Jhansi, India}
 \country{India}}

\author{Chandresh Kumar Maurya}
\email{chandresh@iiti.ac.in}
\affiliation{%
 \institution{IIT Indore}
 \country{India}}

\author{Shitala Prasad}
\email{shitalaprsd@gmail.com}
\affiliation{%
 \institution{Institute for Infocomm Research, A*Star}
 \country{ Singapore}}





\begin{abstract}
 Intelligent recommendation and reminder systems are the need of the fast-pacing life. Current intelligent systems such as Siri, Google Assistant, Microsoft Cortona, etc., have limited capability. For example, if you want to wake up at 6 am because you have an upcoming trip, you have to set the alarm manually. Besides, these systems do not recommend or remind what else to carry, such as carrying an umbrella during a likely rain.  The present work proposes a system that takes an email as input and returns a recommendation-cum-reminder list.  As a first step, we parse the emails, recognize the entities using named entity recognition (NER).   In the second step, information retrieval over the web is done to identify nearby places, climatic conditions, etc. Imperative sentences from the reviews of all places are extracted and passed to the object extraction module. The main challenge lies in extracting the objects (items) of interest from the review. To solve it, a modified Machine Reading Comprehension-NER (MRC-NER) model is trained to tag objects of interest by formulating annotation rules as a query. The objects so found are recommended to the user one day in advance. The final reminder list of objects is pruned by our proposed model for \emph{tracking objects} kept during the "packing activity." Eventually, when the user leaves for the event/trip, an alert is sent containing the reminding list items. Our approach achieves superior performance compared to several baselines by as much as {\bf 30\% on recall} and {\bf 10\% on precision}.
\end{abstract}

\maketitle



\section{Introduction}

We often forget things to carry at the last moment when we are leaving for office, some tourist place, or as simple as forgetting notes given in an email invite.   Current reminder systems are very limited in capability. They work only in certain situations. For example, reminding about when to take medicines. These systems make use of a predefined list of activities or events and are limited in scope. Some of the existing systems employ activity recognition to remind. Again such systems make limited use of the information available over the web, nor do they work for generic tasks. Reminders systems available in Siri, Google Assistant, Alexa, or any other electronic device work based on the user input and lack cognition or intelligence. For example, if you want to wake up at 6 am because you have an upcoming trip, you have to set the alarm manually.  The reminder system plays a vital role in various strata of life. Their need comes out due to the forgetfulness of the human brain or decline of memory over time or diseases such as Alzheimer's, Parkinson's, etc. Thus reminder systems can help people suffering from any memory issues to take medicines, do daily chores, etc. 

 The proposed system in the current work aims to solve two important issues:
 \begin{enumerate}
     \item (RQ1) what to remind?
     \item (RQ2) when to remind? 
 \end{enumerate}
 The first research question deals with coming up with the right reminder list curated intelligently for the problem at hand. The second research question deals with the problem of sending push notifications to the user. For example, reminding the user to carry the umbrella when the user is already outside is useless. In the present work, we make an effort to answer such research questions. Specifically, we develop a system that takes as input an invitation such as emails or ticket booking information. This information triggers the reminder system to look out for sources where it can gather data for reminding. In particular, it reads user email to understand user itinerary and extract special notes. What to remind based on the environmental conditions such as weather, it crawls the web for things useful in a certain weather condition. For example, carrying an umbrella in bad weather. More details of the proposed system can be found in the system description section. Concretely, we make the following contributions \footnote{code+data will be available at https://github.com/recumremsys}:
\begin{itemize}
    \item A system is proposed to recommend and remind the user of the things that might be useful to him/her at the right time. 
    \item An object detection algorithm is proposed to vet the recommendation list based on the user's packing activity.
    \item Effectiveness of the proposed system is evaluated on the test data from the tourism domain and shows impressive results compared to several baselines.
\end{itemize}
We reiterate that the proposed approach is not \emph{limited} to the travel domain. It applies to other domains such as birthday parties, wedding or medicine reminders, etc., depending on what information is present in the emails and corresponding reviews availability.  Another concern we would like to address is that our proposed system (which is likely to be released as an app) asks permission to access user's emails, similar to google calendar and other apps in the mobile devices.  To solve the privacy concern, before sending the server's location and date, we ask the user to provide consent. 

The rest of the paper is organized as follows. The related work is presented in section \ref{related} followed by the proposed system architecture in section \ref{proposed}. In section \ref{training}, we discuss the testbed and setup for training the modules followed by baselines in section \ref{baselines} where we compare our work with various known information retrieval techniques. Finally, section \ref{conclusion} concludes the work with future directions.
 
\section{Related Work} \label{related}
There is a plethora of works on the reminder system. Most of these reminder systems are either time or location-based. Time-based reminder systems work on the principle of finishing the tasks in a To-do list within a deadline. On the other hand, a location-based reminder system uses GPS or google map data to identify the resolution place where the To-do list tasks can be accomplished. There is another emerging category of reminder systems based on the context. In this section, we provide a summary of the reminder systems based on the above concepts and identify the research gaps.

In \cite{ludford}, the author presents a system based on GPS-location data.  In a similar line, the work in \cite{sohn2005place} reminds the user by taking into account the user's position obtained from the cellphone tower.  In both of these systems, the input is the resolution place where the reminder message will be prompted. The author of \cite{tanaka2016advancement} uses indoor positioning and movement state to identify resolution place.  Along similar lines, the work in \cite{suzumura2018can} expects a To-do list for resolution place. They build a classifier using To-do as features and resolution place as class labels. Very recent work on location-based reminders is called SmartNotify \cite{li2018smartnotify}. It updates the user's preference list based on her activity and provides a suggestion for point of interest (POI) using clustering and association rule mining. Recommendations through Google maps are also location-based and are the same for every user \cite{pamulapati2017idoremind}. RemindMe \cite{ertugrul2014remindme} supports geo-tagging via google maps and foursquare for location-based reminding. In \cite{graus2016analyzing}, Graus et al. present an analysis of a time-based reminder using logs produced by Cortana (a Microsoft Inc. automated personal assistant like Siri, Alexa, etc.). They identify common tasks across a range of users and temporal patterns of setting reminders. They also conclude that reminder setting time is a powerful feature for reminder notification time.

In \cite{kim2004gate}, the author proposes an appliance in a family shared home located at the front door area that represents informative messages, reminding users of things they need to take and know before leaving home. 
In \cite{williamson2013designing}, introduces  designing interactive system for elderly people in the form of a smart pen and paper calendar-based reminder system for the home, which focuses a link between paper interaction and scheduling reminders, such as using smart pen annotations and using the location of written reminders within a paper diary to schedule digital reminders. The results also describe important physical aspects of paper diaries discussed by older adults, such as daily/weekly layouts and binding.
In \cite{abdul2012elderly}, the author proposes facilities a reminder system should consist for older people by conducting a series of studies: interview, usability evaluation, and drawing activity, which resulted in revealing that elderly users expected that a reminder system should be simple, familiar, flexible and recognizable to them. 

In \cite{nate2016smart}, the author proposed a system to remind three kinds of things: Communication reminder for calls and messages, a basic reminder for schedules and medicines, and location-based reminders for POI. iDoRemind system \cite{pamulapati2017idoremind} allows people to create reminders based on location and time restrictions, suggest locations based on time and reminders.
Both of the above systems rely on manually creating a To-do list and do not automatically update the user's preference. 

Another body of work for reminder systems is based on the context. In \cite{gaurav2015offline}, the author proposes an offline strategy to collect context in the form of stored messages, emails, logs, chat history, call history, and events such as birthdays. However, they do a very high-level analysis of emails, messages, etc., for composing reminders such as reminding missed class, unread emails, and so on. In our work, we go one step further to understand emails, chats to look for special notes, travel itineraries, and so on using natural language processing (NLP). In \cite{watanabe2019investigation}, the author investigates the user's context based on her physical activity, such as sending emails when the user is about to reach a destination. In \cite{hong2009context}, Hong et al. propose an agent-based framework using context-history for providing personalized services. They intelligently mix the user's profile and context history for extracting useful patterns.

\begin{figure}[ht]
\small
    \centering
    \includegraphics[width = 5cm]{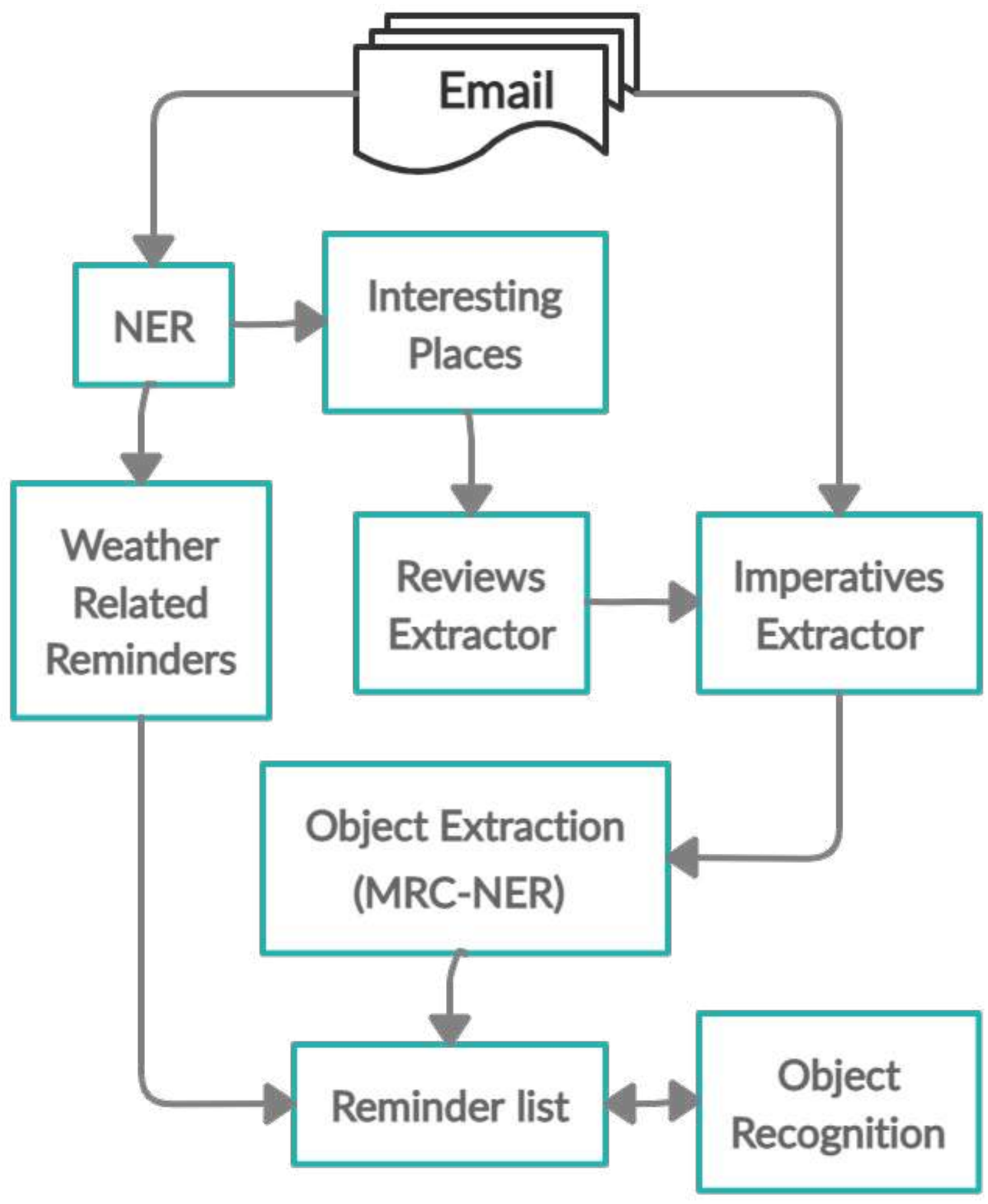}
    \caption{The proposed system architecture}
    \label{fig:system}
\end{figure}

\section{The proposed System} \label{proposed}
In this section, we elucidate the proposed system and its various components. The proposed architecture of the system is shown in Figure \ref{fig:system}. The user interacts with the system by feeding in a raw document (e.g., an Email, SMS, etc.). To show the proof of concept, we develop our system, which sends reminders in the \emph{Tourism} domain. That is, emails/SMS are related to travel. The named entity recognition (NER) module extracts the named entities present in the email. The PLACE Named entities found in the previous step are used to crawl sites like tripadvisor.com for places-of-interests (POI) around the PLACE entity. We then crawl reviews related to each POI from websites such as \emph{Tripadvisor, Google Reviews}. Each review is then parsed, and \emph{imperative sentences} are extracted. Such sentences might contain information that should be reminded. These imperatives are used in the downstream task for building a To-do list. We train an MRC-NER model that can tag reminder objects in a sentence. This list is recommended to the user and the user selects some or all of the items to be reminded on the day of the trip. The weather-related reminder module reminds a list of items to carry in particular weather, and this list is clubbed with the list obtained by the MRC-NER module. Finally, we perform object detection and recognition during the user's packing activity for validating the items in the To-do list so that we do not remind items that the user has already packed. In the upcoming sections, we describe each component of the system in detail.

\subsection{Named Entity Recognition}
The proposed system takes email/SMS as input, which is passed through the NER model. We use Stanford's pre-trained Named Entity recognition model \cite{finkel2005incorporating}, which can tag 7 classes: LOCATION, PERSON, ORGANIZATION, MONEY, PERCENT, AND DATE in a raw text. Interesting entities for our use case are LOCATION and DATE. The LOCATION entity is used as a resolution place to search for POI via Google's Search API \footnote{https://pypi.org/project/google/} and Tripadvisor \footnote{https://www.tripadvisor.com/}. The DATE is used for weather-related recommendations and as a trigger point when pushing the user's reminder list (for solving RQ2).

\subsection{Places-of-Interests (PoI)}
Names of interesting places on which the user can do interesting activities like trekking, beach walking, etc., are extracted by firing a query on Google search API. The query is \quotes{things to do in} + [LOCATION] + \quotes{tripadvisor.com}. From the search results, the URLs with the term \quotes{Tripadvisor} are filtered out. By scraping search results, we build a list of interesting POIs.

\subsection{Weather Related Reminders}
The arrival date, departure date (if available), destination place is passed as input to DARK SKY API \footnote{https://darksky.net/dev} for searching the weather conditions of the destination place from the duration the user arrives at the destination and departs from there.  If the departure date is not mentioned in the document, we assume it to be a week later. We recommend certain items to the user from the API results using the following conditions. If any of the conditions holds, its objects are added to the user's recommendation list based on the common experience: (a) {\bf Minimum Temperature:}
If the temperature is predicted to be less than 30$^\circ$F during the time of visit, then \quotes{boots, thick sweater, a winter coat}. are recommended. If it is less than 50$^\circ$F and more than 30$^\circ$F, then \quotes{light sweaters, long pants, gloves, hats, acrylic fiber clothes} are recommended, (b) {\bf Maximum Temperature:}  If the temperature is predicted to be more than 50$^\circ$F and less than 70$^\circ$F during the time of visit, then \quotes{shorts, t-shirt, water bottle} are recommended, and if it is more than 70$^\circ$F then \quotes{light-colored dress, cotton clothes, water bottle} are recommended, (c) {\bf Percentage of rain:} If the percentage of rain is more than 40\%, then \quotes{Ankle boot, Umbrella, Raincoat} is recommended.

\subsection{Reminder from Reviews}
From our analysis, we find that people usually suggest  reminder items on travel websites such as Tripadvisor \footnote{https://www.tripadvisor.in/} in reviews. However, most of them might go waste as users are hardly reading them due to the explosion of big data \footnote{https://tinyurl.com/2p62yevj}. As such, users feel lost while looking for relevant and meaningful information from travel websites. Therefore, the authors propose a pipeline to extract these hidden reminders from the  reviews about POIs collected from Tripadvisor website.
\subsubsection{\bf Imperatives Extraction} \label{imperative}
We hypothesize that  reminder items are hidden inside of an imperative sentence. For example, \emph{\quotes{Don't forget to take umbrella as it might rain}}. We validate our hypothesis in the experiment section. Imperative sentences have either one of these two properties. They start with a verb, or  They end with a "!".
Each review is tagged using Stanford's corenlp \cite{conf/acl/ManningSBFBM14}, and sentences satisfying the above two conditions are retrieved. 

\subsubsection{Machine Reading Comprehension - Named Entity Recognition (MRC-NER)}
The extracted imperatives contain useful reminder information.  It is observed that the entity found by the standard NER module is not enough. For example, \emph{``don't try to take a dip in the \quotes{water}, many have died here} and \emph{``don't forget to bring \quotes{water} with you as it gets really hot}, in both the sentences, \quotes{water} is considered as an entity by the standard NER model. But, in reality, \quotes{water} is being suggested only in the later review.  The context of the review plays a significant role in deciding whether to suggest the entity or not. Considering this, the authors chose to use the Machine Reading Comprehension-based Named Entity Recognition (MRC-NER) \cite{li2019unified} approach. This approach reformulates a standard NER task into QnA task.  We observe that there can be no case of nested entities in this task. Hence, we modify the model's architecture and use only the entity's start and end predictions. Training details of MRC-NER are given in section \ref{training}.

\subsection{Object Recognition Module}
Items to be reminded of (also called objects in the previous section) may or may not be useful depending upon the user's interest. In other words, objects found in the previous steps are recommended to the user one day before the start of the journey. (S)he can select which items need to be reminded. A pruned list of items is then kept in the reminder list. This list is not useful unless it can be pushed to the user at the right time after verification. In a way, we want to simulate the behavior of what our elders/parents remind us when we are leaving for a tour. In the next step, we want to validate whether the user has packed the items or not. Therefore, we build a model for recognizing objects kept in the travel bag during the packing activity. The model works under the assumption that we are packing one item at a time in front of a smart surveillance camera. There may be questions that packing one item at a time is not realistic but we claim that no existing machine learning model can detect items when wrapped together for example cloths. We verified detecting multiple items together but accuracy was poor. The assumption of packing one item at a time is reasonable for some items like \emph{water bottle, wallet, umbrella etc}. Hence, to make the system work, we have to rely on assumption stated earlier. Secondly, packing in front of a camera seems unrealistic but we believe will become prevalent in future when more and more people start living in \emph{smart homes}. So in a way, our proposed system falls in the cateogry of \emph{futuristic technology}.

Our goal is to detect the object packed by the user from live video feeds. Object detection module consists of three models: U2net \cite{u2net}, YOLOv3 \cite{redmon2018yolov3}, and Stand-Alone Self-Attention model (SASA)\cite{ramachandran2019stand}.  The pre-trained versions of the YOLOV3 and U2net models are used while SASA is trained on fashion dataset \footnote{https://www.kaggle.com/paramaggarwal/fashion-product-images-dataset}. 

\subsubsection{Datasets:}
The Stand-Alone Self-Attention model is trained on extracting  7000 images from 10 categories from the fashion dataset.  Since there is a class imbalance present for all classes in the dataset, image augmentation techniques such as flip, rotation, shearing, cropping are used to counterbalance them thereby increasing the number of images per class. After the data augmentation, the total number of images is 14000.

\subsubsection{\bf The Object Detection Algorithm}
\begin{figure}
\centering
  \subfloat[]{\includegraphics[width = 2.7cm,height = 2.7cm]{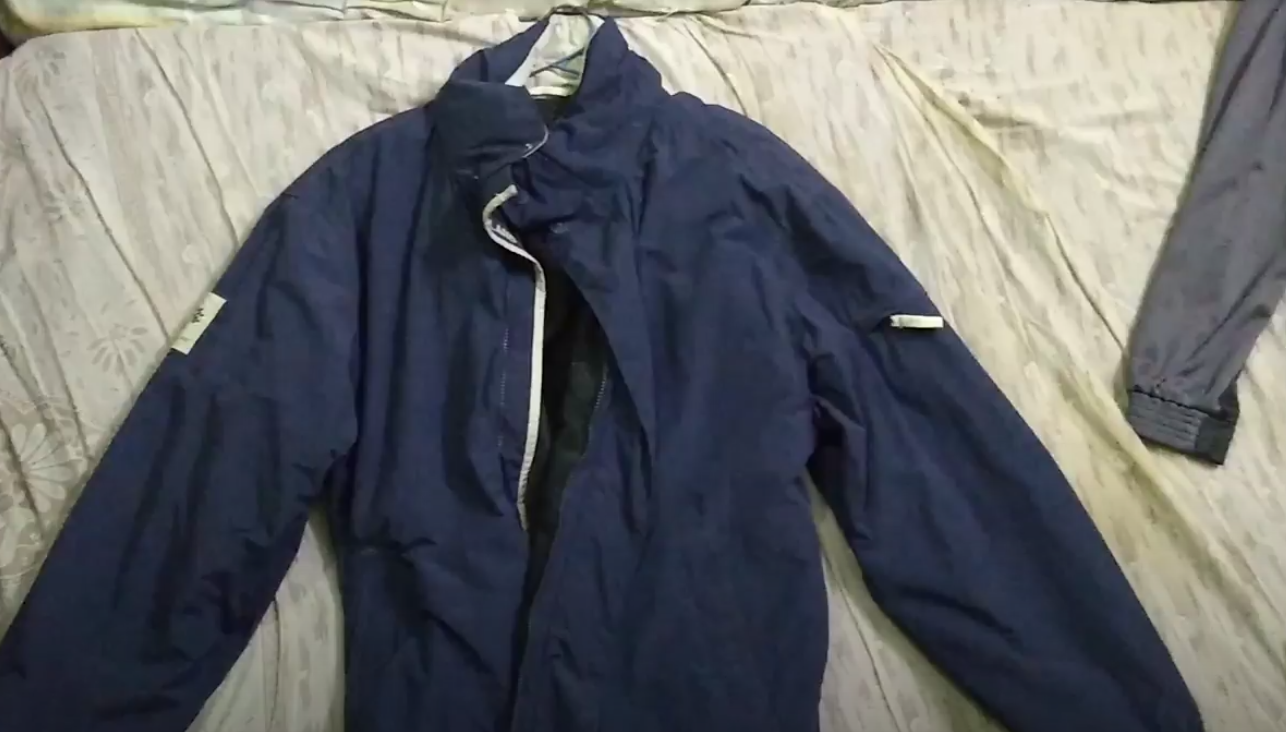}}
  \subfloat[]{\includegraphics[width = 2.7cm,height = 2.7cm]{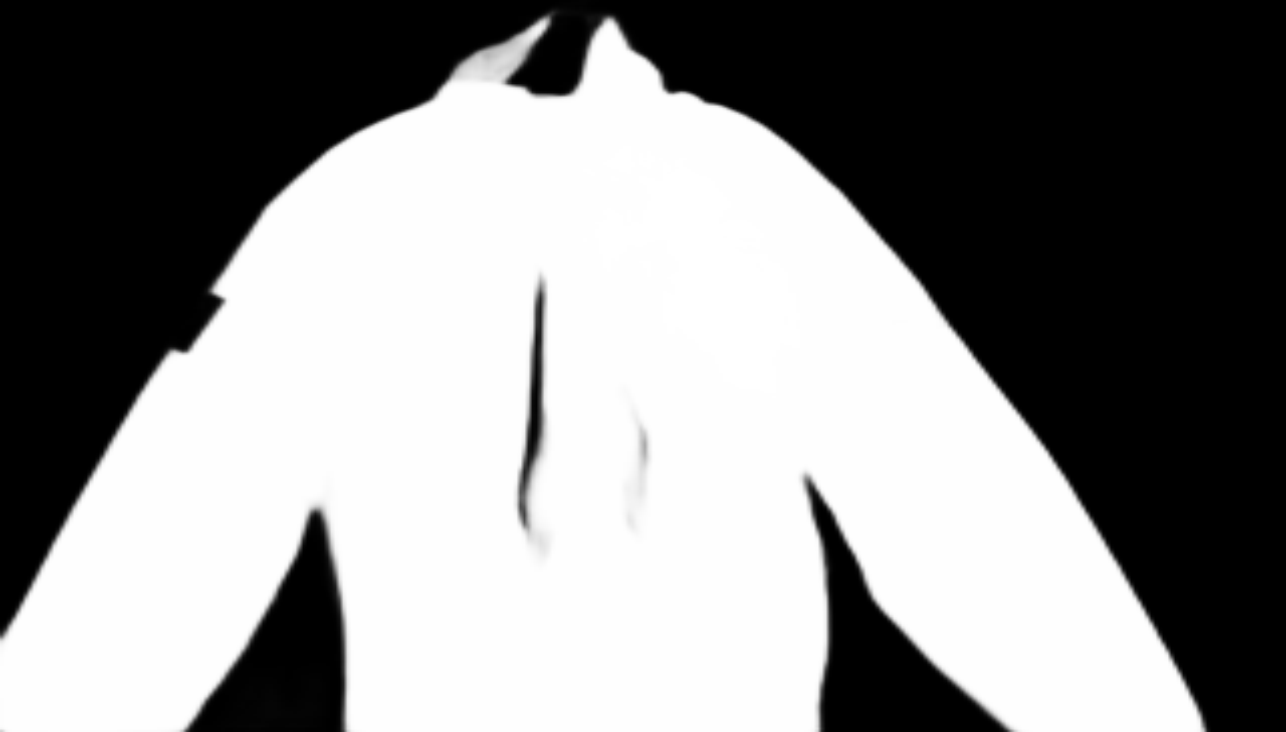}}
  \subfloat[]{ \includegraphics[width = 2.7cm,height = 2.7cm]{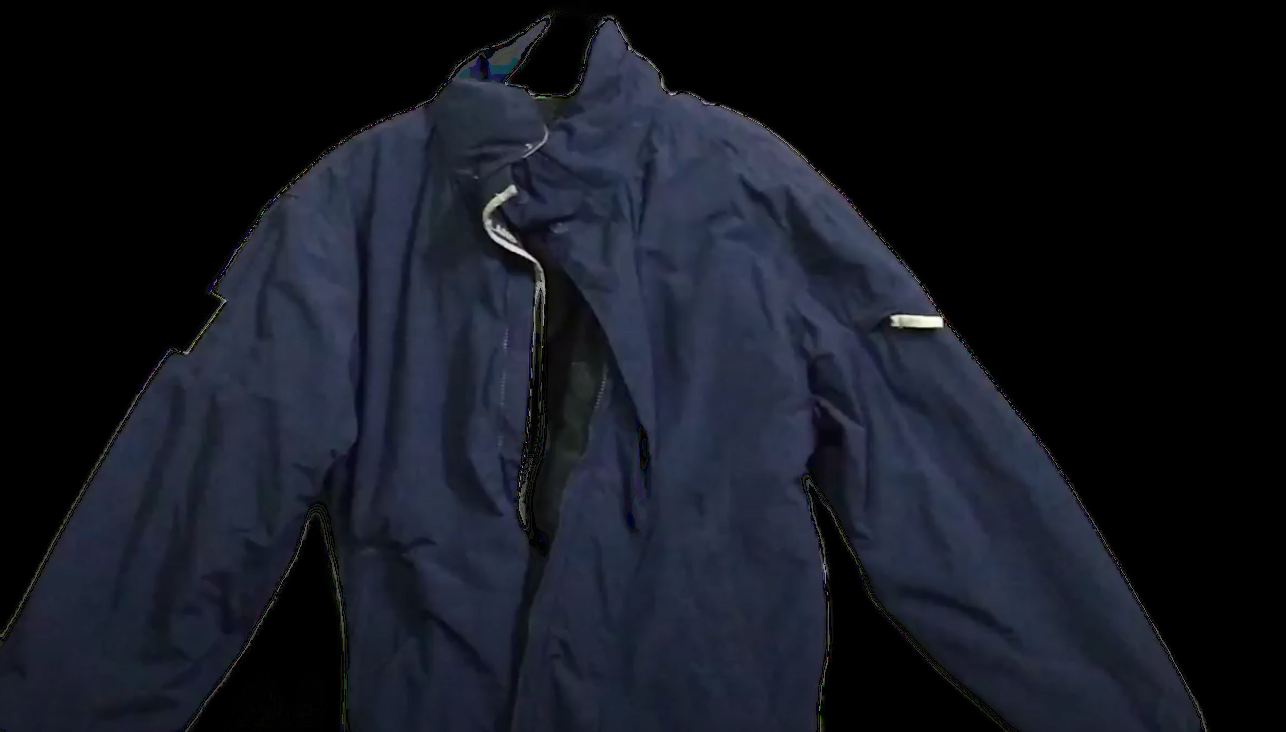}}
\caption{(a) Sample Image (b) Mask of Sample Image(Extracting Salient Object using U2net), (c) Masked Sample Image(using mask shown in image (b))}
\label{u2netoutput}
\end{figure}

The object detection algorithm is shown in Algorithm \ref{objdetect}.  Before applying any model, the variance of the  Laplace transform of each frame is computed for detecting \quotes{blur} in the image. The reason is that users can pack items in a hurry, and items may be in motion and difficult to identify. The presence of the blur negatively affects the performance of the models used. Therefore, frames whose \quotes{blur} is below a threshold $\gamma$ (=40 as found in the test videos) are rejected for further processing. Subsequently, the frame is passed to YOLO for object detection. If the prediction score is more than a threshold $\delta=0.70$ (a user defined parameter) and the object belongs to the COCO data category, it is retained otherwise passed to the U2net model to extract the salient object present in the image shown in fig. \ref{u2netoutput}(b).   Directly using the salient object for detection suffers from poor performance as most of the information like texture and other important features are lost. Therefore we mix the salient object with the real-world image and convert it into a color image, as shown in fig. \ref{u2netoutput}(c). Finally, the mixed frame is passed to SASA for object detection. 

\begin{algorithm}
\small
\SetAlgoLined
\KwIn{ $F \rightarrow$ frames in the video, $L \rightarrow$ list of objects, 
$yolov3obj \rightarrow$ objects detected using YOLOV3,
parameter$\rightarrow \gamma, \delta$}
\KwResult{$reminders$}
\SetKwInOut{Input}{Initialize}
\Input{ $count = 0, reminders =[], \gamma = 40, \delta=0.7, window = 15 $}
 \For{$f$ in $F$}
 {
 $blur\_amount = Laplace\_variance (f)$ \;
 \If{$blur\_amount > \gamma$}
 {
  $pred, score = YOLOV3(f)$\; 
  \lIf{$pred \in yolov3obj$ \&\& $score >\delta$}
    { $reminders = reminders + pred$}
 \Else{
  $framenew = U2net(f)$\;
  $framemixed = framenew + f$\; 
 
   $pred = SASA(framemixed)$\;
   $reminders = reminders + pred$\;
  
 }
 }
  }

  \KwRet{reminders}
 \caption{Object Detection Algorithm}
 \label{objdetect}
\end{algorithm}

\section{Training \& Evaluation} \label{training}
In this section, we present the training procedure of MRC-NER,  SASA, followed by the evaluation of the (1) object extraction pipeline, (2) object detection pipeline, and (3) the overall system.

\subsection{Training of MRC-NER}
To prepare the data for training the MRC-NER model, we extracted 10000 imperative sentences from reviews using the same rules described in section \ref{imperative}. The data is divided into three equal parts and assigned each part to a group of 3 annotators. The annotation rule is to label things that a person might want to carry along while going on a trip. The average inter-annotator metric (kappa score) is 0.88.

Formulating NER task as a QnA task required the formulation of an appropriate query. We use the guidelines that are given to the annotators to form a query. For example, one of the queries is: \quotes{\textit{/What are the things a person might want to carry along while going on a trip?}} The answer returned is formed by taking entity spanning from start token to end token. The entity between the start and end token represents the item of interest.
The MRC-NER model is implemented using Keras available in Tensorflow, minimizing the binary cross-entropy loss. The data is split in the ratio of 60:20:20 for train, validation, and test, respectively. The model is trained for 100 epochs and achieved 89\%/83\%/84\% F1 score on train/validation/test sets, respectively. 
\subsection{Training of Stand-Alone Self-Attention model}
The Stand-Alone Self-Attention model is implemented using Keras in TensorFlow. The architecture is made using a padding layer, followed by a convolutional layer, followed by  3 Identity blocks. 
The first block has an attention stem \cite{ramachandran2019stand} in place of the spatial convolution between an upsampling and downsampling convolution layer with batch normalization after each layer. Simultaneously, the other two contain the same architecture, with the attention stem replaced by the attention layer. The convolution data is split into train and test in the ratio of 80:20 using StratifiedKFold, and the model is trained for 10 epochs. It achieves an accuracy of 94.26\% on the training set and 95.43\% on the validation set.

\subsection{Evaluation of object extraction pipeline}
 Objects to be recommended and reminded are extracted from the POI's review using the imperatives extraction module followed by the MRC-NER module. To evaluate the pipeline,  500 POIs and their most recent 30 reviews are crawled from Tripadvisor. These reviews are given to adult users and asked to identify objects mentioned in the review, which they would like to carry with them while visiting the POIs. This setting is very realistic as we often look out for TripAdvisor's review while visiting new POIs. Objects extracted by users serve as the ground truth for these POIs. The object list returned by our module is evaluated against the ground truth. The average precision (the fraction of retrieved items that are relevant) and recall (the fraction of relevant items that are retrieved) are 74.38\% and 89.89\%, respectively. The average recall clearly shows that our system is close to human performance in this tedious task.  The relatively low precision is due to some objects which can not be carried, such as {\it beaches}. This module's output is recommended to the user one day in advance (answer to RQ1). The user then asks the system a list of items selected from the recommended list to be reminded on the day of the trip. This pruned list of items is then used as ground truth for the next module in the pipeline.


\subsubsection{\bf Evaluation of the object detection algorithm}
\begin{figure}[t]
\centering
 \subfloat[]{ \includegraphics[width=2.7cm, height=2.7cm]{images/final_img.pdf}} 
  \subfloat[]{\includegraphics[width=2.7cm, height=2.7cm]{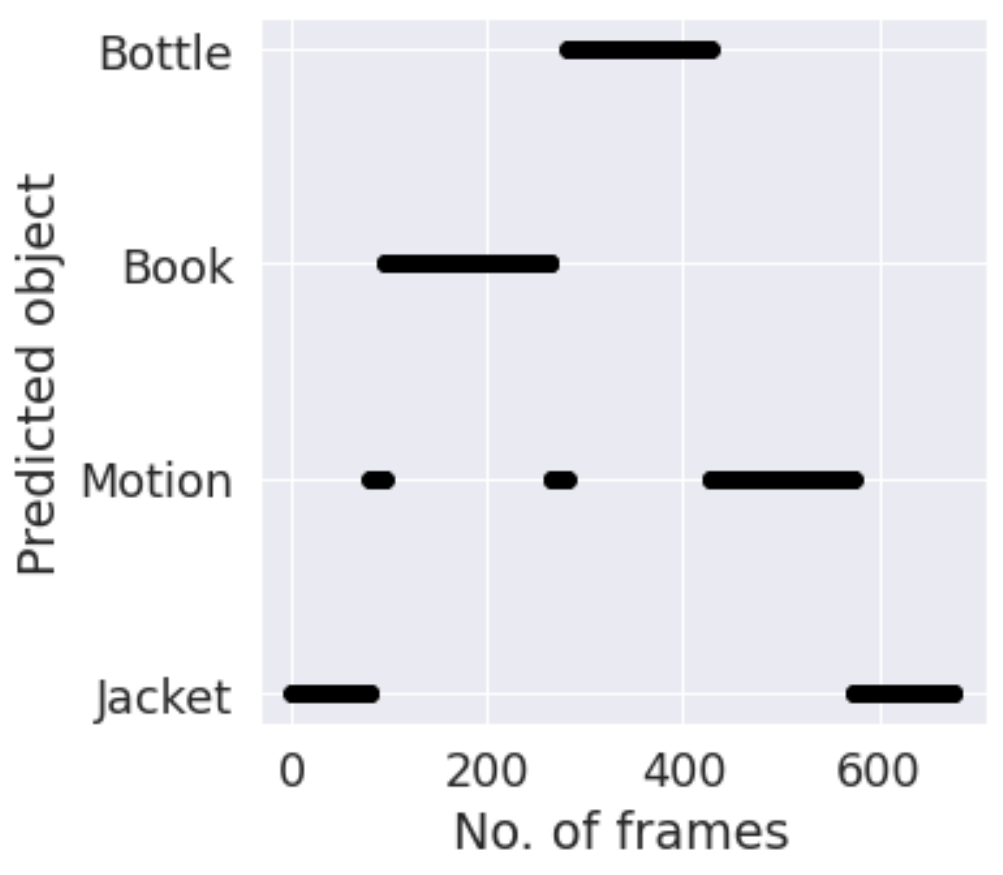}}
  \subfloat[]{ \includegraphics[width=2.7cm, height=2.7cm]{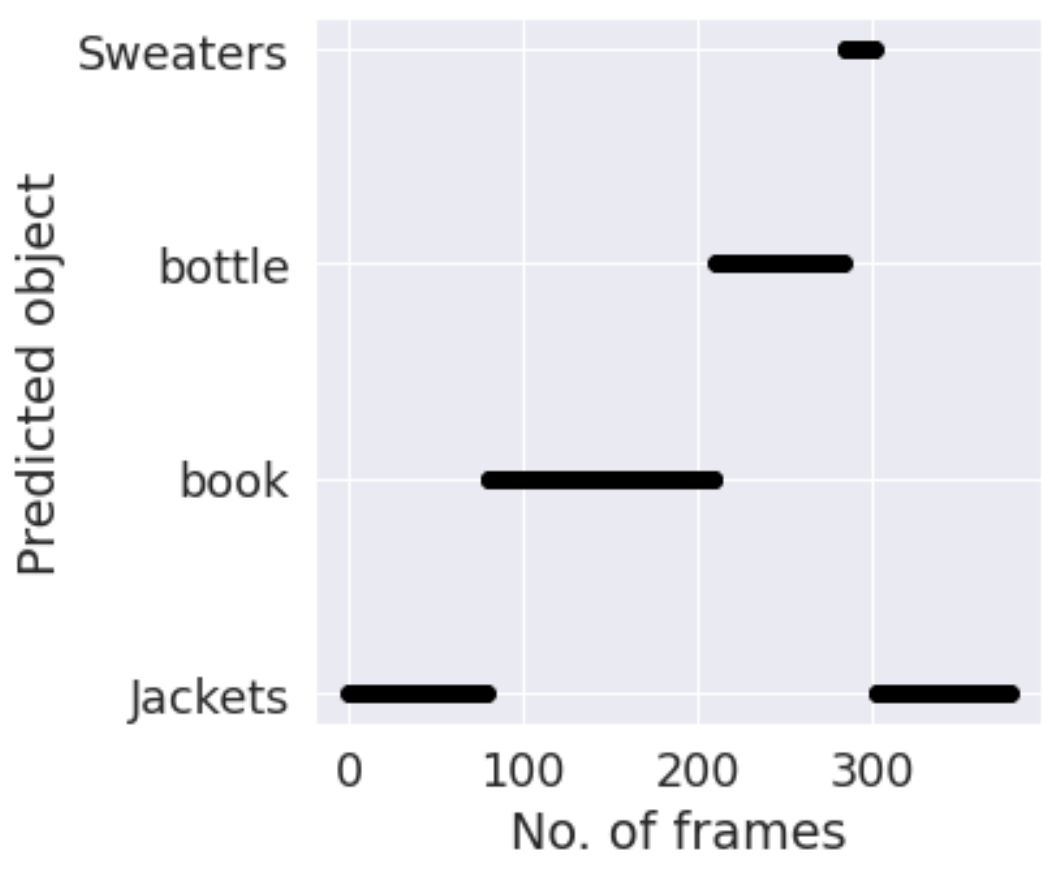}}
 
\caption{Performative analysis of object detection module: (a) Objects present in the video, (b) Predictions per frame without blur and smoothing operation, (c) Predictions per frame with blur detection and smoothing\protect\footnotemark}.
\label{original}
\end{figure}

\begin{figure*}
    \centering
    { \includegraphics[width=13cm]{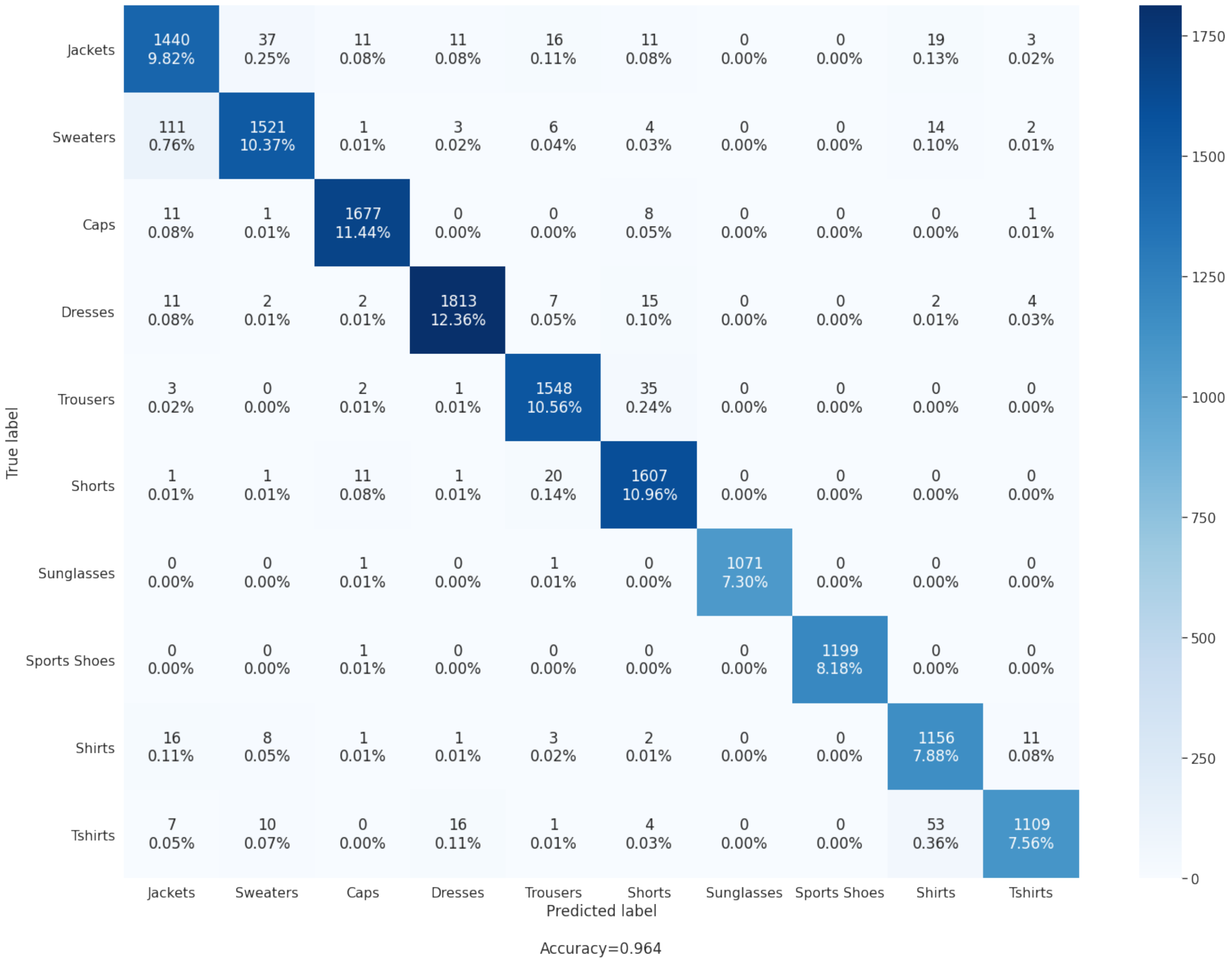}}
    \caption{Category-wise confusion matrix of object detection module for validation}
    \label{confusion}
\end{figure*}

In this section, we first show the effect of blur detection and smoothing operation. The results are shown in fig. \ref{original}. There are four objects present in the ground truth video, as shown in fig. \ref{original}(a). The object detection module's result without blur detection and smoothing as shown in \ref{original}(b) generates lots of objects in different frames (motion is not an object). That means it flags lots of false positives. To remedy this, we apply the blur detection available in the OpenCV \cite{opencv} on frames and subsequently remove the frames having blur amount lower than a threshold $\gamma=40$ (found via hyperparameter search). As a result, the number of false positives goes down. The object detection module's accuracy is further improved by the smoothing operation, as shown in fig. \ref{original}(c).  This concludes that blur detection with smoothing operation is an effective approach. In fig. \ref{confusion}, the confusion matrix shows the object detection module's overall performance. We can see in the confusion matrix that the more similar looking objects such as sweaters and jackets are difficult to classify.
\footnotetext{Number of frames in the Figure 3(c) is less because of the removal of the blur frames}

The object detection module is used to track the items in the user's packing activity. For the evaluation of this module, we ask users to make videos of the items they wish to carry on their trip. We ask them to analyze the reviews of 500 POIs to evaluate the object extraction module. The videos are then passed through the object detection module, and predictions are compared with the user's ground truth (pruned list). The accuracy of the module on the test videos is 82.58\% . Upon investigation,  the reason for the module's low performance is found to be the presence of UNKNOWN objects (objects the module is not trained to detect). In the future, we plan to train the SASA model to detect more categories. 
\subsubsection{Evaluation of the overall system}
 To evaluate the system as a whole, we pass 500 emails as input to the system. The output of various modules is shown in Table \ref{email1}.  From the result, we can observe that the NER module can extract places like \quotes{\textbf{Newport}}, person like \quotes{\textbf{Cecil Dawson}}, etc. POI output comes from the Tripadvisor website. The weather recommendation module's output suggests the user carry \quotes{\textbf{light sweaters, long pants, gloves, hats, acrylic fiber clothes}} since Newport is  \quotes{\textbf{chill in winter}}. The MRC-NER suggests user carry \textbf{\quotes{ID, card, money, shoes, credit card, food, water}}, etc. These recommendations are presented to the user one day before the trip start day. The user selects the item (s)he wishes to carry. Finally, these items are tracked via the object detection module and pushed to the user 1 hour before the start of the trip (ans. to RQ2). The result is shown in the last row of the Table \ref{email1}. We can see that the final reminder list clearly detects important items for the trip such as \textbf{\quotes{water, jacket, hat}} which might be useful while roaming near Newport.

\begin{table*}
\caption{Output of the system on a sample email}
\label{email1}
\begin{tabular}{|p{2cm}|p{13cm}|}
\hline
Document                            & To, Drake Wingly, 1722 Lincoln Drive Rose Park, FL 07662 Subject: Invitation to Motivate Tech Growth Conference Dear Mr. Wingly, As a representative of Motivate Tech Growth Conference, I am pleased to invite you to our inaugural technology conference that will be taking place on Nov 30, 2020. This conference brings together the 5 top Technology firms in the country to bring in light the best of Technological nerds for some discussions on the direction and growth of technology for the nation and the world in the upcoming two decades. We would be thrilled to have you present at this conference and to hear from you about a few new the technology advancements and their impact on different business markets and daily lives. We would also love to hear your thoughts and opinions in this direction. Please respond to our invitation to you before Nov 25, 2020, to secure a place before passes are open to the public by Nov 28, 2020. We look forward to your positive response to the Motivate Tech GrowthConference. Don't forget to bring your student \hl{ID Card}. Note: Newport is quite chill in winter, don't forget to bring \hl{jacket}. Regards, Cecil Dawson Conference Representative, Motivate Tech Growth Conference. \\ \hline
NER                                 & \{'DATE': {[}'Nov', '20', ',', '2020', '1722', 'Nov', '30', ',', '2020', 'Nov', '25', ',', '2020', 'Nov', '28', ',', '2020', 'winter'{]}, 'LOCATION': {[}'Newport', 'FL', 'Newport'{]}, 'ORGANIZATION': {[}'Motivate', 'Tech', 'Growth', 'Conference', 'Motivate', 'Tech', 'Growth', 'Conference', 'Dear', 'Motivate', 'Tech', 'Growth', 'Conference', 'Motivate', 'Tech', 'Motivate', 'Tech', 'Growth', 'Conference'{]}, 'PERSON': {[}'Cecil', 'Dawson', 'Drake', 'Wingly', 'Wingly', 'Cecil', 'Dawson'{]}\}                                                                                                                                                                                                                                                                                                                                                                                                                                                                                                                                                                                                                                                                                                                                             \\ \hline
POI                                 & 1. Yaquina Head Outstanding Natural Area 2. Devils Punchbowl State Natural Area 3. Oregon Coast Aquarium 4. Yaquina Bay Lighthouse 5. Newport's Historic Bayfront 6. Hatfield Marine Science Center 7. Nye Beach 8. Cape Foulweather 9. Yaquina Bay Bridge 10. South Beach State Park                                                                                                                                                                                                                                                                                                                                                                                                                                                                                                                                                                                                                                                                                                                                                                                                                                                                                                                                                                     \\ \hline
Weather                             & {[}('2020-11-25 Wednesday', '35.16°F', '42.6°F', 'Partly cloudy throughout the day.', 17.0), ('2020-11-26 Thursday', '34.79°F', '42.22°F', 'Partly cloudy throughout the day.', 17.0), ('2020-11-27 Friday', '34.42°F', '41.85°F', 'Partly cloudy throughout the day.', 17.0){]}                                                                                                                                                                                                                                                                                                                                                                                                                                                                                                                                                                                                        \\ \hline
weather recommendation & 'light sweaters, long pants, gloves, hats, acrylic fibre clothes'                                                                                                                                                                                                                                                                                                                                                                                                                                                                                                                                                                                                                                                                                                                                                                                                                                                                                                                                                                                                                                                                                                                                                                \\ \hline
MRC-NER & {[} \hl{'ID'}, \hl{'Card'}, \hl{'Jacket'} ,'money', 'day pass', 'ticket', 'walking shoes', 'ticket', 'water', 'pictures', 'fee', 'hat', 'shoes', 'tour buses', 'credit cards', 'comfortable clothes', 'food', 'VIP pass', 'Bag', 'Book', 'Laptop'{]}                                                                                                                                                                                                                                                                                                                                                                                                                                                                                                                                                                                                                                                                                                                                                                                                                                                                                                                                                                                                                                                             \\ \hline

Object Detection Module Output  & {[} 'Jacket' , 'walking shoes', 'water', 'hat', 'comfortable clothes', 'light sweaters, long pants, acrylic fibre clothes', 'Bag', 'Book', 'Laptop' {]}

                \\ \hline
\end{tabular}
\end{table*}

\section{Baselines}\label{baselines}
In the previous section, we discuss the evaluation of the model proposed. In this section, we compare our work with baselines and show that our approach outperforms all the baselines.  In particular, we compare against LDA \cite{blei2003latent}, TF-IDF, QnA \footnote{roberta-base-QnA-squad2-trained available at https://huggingface.co/
}, Spacy NER \cite{spacy2}, and Popular mentions from Tripadvisor. LDA is used as a topic model on reviews, and top-k topics serve as the objects for the reminder. Answers retrieved by QnA model serve as objects.

\subsection{Latent Dirichlet Allocation}
LDA is a well known topic extraction technique, and we use it to find the most important words from the review text. Based on the assumption that words in the same topic are more likely to occur together, it is possible to attribute phrases or keywords to a particular topic.

\subsection{TF-IDF}
In order to find the most frequently used tokens, we used term frequency document frequency method to extract n-grams (upto n=3). 

\subsection{Popular Mentions}
On tripadvisor website, most insightful tokens are displayed before reviews. To compare our work, we extract these mentions and subjected them to the same evaluation procedure. The results are shown in the Table 5. The results are found to be unsatisfactory but with a bit of improvement from the above techniques.

\subsection{Question Answering Model}
The QnA models trained on large datasets like Stanford Question Answering Dataset (SQuAD) \cite{rajpurkar2016squad} generalizes well even for unseen tasks. We use QnA model implemented by Hugging Face
\footnote{https://huggingface.co/}. For fair comparison, we use the same query that we use for training of MRC-NER model.

\subsection{Spacy NER}
The authors train a custom NER model over the NER model in Spacy \cite{spacy2}. It allows training a model over one relevant category with a relatively small number of labeled samples. For fair comparison, the model is trained on the same dataset of 10000 sentences used to train the MRC-NER model.

\subsection{Results of Baselines}
The results are shown in Table \ref{Baselines}. The reason for poor results with unsupervised information retrieval techniques (TF-IDF) is the less likelihood of finding the reminders in a vast amount of reviews. The results from popular mentions are better but still not satisfactory. The QnA model trained on SQuAD outperforms the previous techniques by a large margin. Still, the model's relatively low performance is because of the presence of multiple entities in a single review and the complexity of this task.
Spacy NER outperforms even the pre-trained QnA model but still fails to generalize well because of the complexity of the task. We can see that our approach outperforms all the previously mentioned techniques in precision and recall.
The qualitative results are shown in table \ref{Baselines Results}. The  \quotes{Review} column contains respective reviews, and other columns contain output from each approach. The ground truth is shown in bold and a \quotes{--} means the method does not return anything.  We can clearly see that our approach can extract matching ground truth objects. However, sometimes our approach adds false positives such as \quotes{flip flop} in the last row.
\begin{table*}
\caption{Baselines}
\label{Baselines}
\centering
 \begin{tabular}{|c| c| c|}
 \hline
 Approach & Precision & Recall \\ [0.5ex] 
 \hline\hline
 LDA & 2.049\% & 2.350\% \\
 \hline
 Tf-idf & 2.350\% & 2.741\%\\
 \hline
 Popular Mentions & 8.524\%& 7.679\%\\ 
 \hline
 QnA & 52.00\% & 44.45\% \\
 \hline
 Spacy NER & 64.91\% & 59.58\% \\
 \hline
 Our approach & {\bf 74.38\%} &{\bf 89.89\%} \\ [1ex]
 \hline
\end{tabular}
\end{table*}

\begin{table*}
\caption{Qualitative Evaluations}
\label{Baselines Results}
\centering
\small
 \begin{tabular}{|p{2.8cm}| p{1.3cm}| p{1.3cm}| p{1.3cm}| p{1.3cm}| p{1.3cm}| p{1.3cm}| p{1.3cm} |}
 \hline
 Review & LDA & Tf-idf & Popular Mentions & QnA & Spacy NER & Our Approach \\ [0.5ex] 
 \hline\hline
  wear proper \textbf{shoes hat water}.  & water & water & water & shoes & shoes, hat, water & shoes, hat, water \\
 \hline
  bring \textbf{water hat umbrella} as it was so so hot  & hat & water & --- & water, hat, umbrella & water, hat, umbrella & water, hat, umbrella\\
 \hline
 take off your \textbf{shoes} to walk on the uneven floors for a bit  they shouldnt complain since the artist makes a big deal about this.  & --- & --- & --- & shoes & shoes & shoes \\
 \hline
 pick pocket warnings all over the place  & --- & --- & --- & pick pocket & --- & --- \\
 \hline
 don't try to take a dip in the water, many have died here. & --- & --- & --- & --- & water & ---\\
 \hline
 be sure to apply \textbf{sun screen} wear a \textbf{hat} and good \textbf{shoes} not flip flop & hat & --- & shoes & sun screen, hat  & sun screen, hat, shoes, flip flops & sun screen, hat, shoes, flip flops\\
\hline 
\end{tabular}
\end{table*}
\section{Conclusion}\label{conclusion}
 A system for recommendation-cum-reminder system is proposed to take email as an input. The input is then passed through various modules such as weather information, object extraction from reviews, and object detection module. Our system recommends a list of items that may be useful to the user while visiting a POI. The user then asks the system to remind it on the travel date. The list is then used as ground truth for tracking objects in the packing activity. After the user finishes the packing, the system sends a notification containing a list of items missed during packing. Our system can be used to send reminder in other situations such as birthday/wedding events and with a little effort. Comparison with baselines shows that extracting items to remind is a non-trivial task and involves intelligent integration of various components. As a future enhancement to the proposed system, an additional module can be attached to retrieves top stories from news and social media to find other things to be reminded of. For example, our system will retrieve information from the web, such as washing hands, carrying masks and hand sanitizers during a recent corona outbreak, and reminding the user while leaving home.
 
\begin{table*}
\caption{Output of the system on a sample email}
\label{email2}
\begin{tabular}{|p{2cm}|p{13cm}|}
\hline
Document                            & To: roger@rmail.com Subject: Invitation letter for the post of Senior Content Writer Dear Mr. James, It gives me immense pleasure to inform you that we have selected you for the post of Senior Content Writer in the New York Firm. The selection has been done based on your educational qualification, work experience and your performance in the interview conducted by our company officials. You did well in the oral interview round and also scored good marks in the written test. Considering all these aspects, our management feels that you would be able to handle the position well. We have also gone through the documents produced by you at the time of interview and verified their authenticity. The details about your salary and remunerations have been mentioned in the sheet enclosed. You are requested to join the company on the November 5. Please be there at the office by 10 am. Do bring your \hl{ID} \hl{card} and \hl{passport}. Thanking you Yours' Sincerely, George Peterson Manager HR \\ \hline
NER                                 & \{'DATE': ['November', '5'],
 'LOCATION': ['New', 'York'],
 'ORGANIZATION': ['Senior', 'Content', 'Writer'],
 'PERSON': ['James', 'George', 'Peterson']\}                                                                                                                                                                                                                                                                                                                                                                                                                                                                                                                                                                                                                                                                                                                       \\ \hline
POI                                 & 1. The National 9/11 Memorial \& Museum 2. The Metropolitan Museum of Art 3. Central Park 4. Empire State Building 5. Top of the Rock 6. Statue of Liberty 7. Brooklyn Bridge 8. Manhattan Skyline 9. The High Line 10. One World Observatory                                                                                                                                                                                                                                                                                                                                                                                                                                                                                                                                                                                                                                                                                                                                                                                                                                                                                                                                                                   \\ \hline
Weather                             & {[}('2020-10-05 Monday', '67.18°F', '78.14°F', 'Overcast throughout the day.', 4.0), ('2020-10-06 Tuesday', '63.48°F', '75.17'), ('2020-10-07 Wednesday', '58.48°F', '73.62°F', 'Rain starting in the afternoon.', 85.0){]}                                                                                                                                                                                                                                                                                                                                                                                                                                                                                                                                                                                                    \\ \hline
weather recommendation & 'Ankle boot', 'Umbrella', 'Raincoat', ' light-colored dress', 'cotton clothes','water bottle'                                                                                                                                                                                                                                                                                                                                                                                                                                                                                                                                                                                                                                                                                                                                                                                                                                                                                                                                                                                                                                                                                                                                                           \\ \hline
MRC-NER  & {[} \hl{'ID'}, \hl{'card'}, \hl{'passport'}, 'food', 'money', 'camera','day pass', 'ticket', 'walking shoes', 'tissues', 'tear', 'wand', 'delicious food', 'toothbrush', 'Binoculars' {]}                                                                                                                                                                                                                                                                                                                                                                                                                                                                                                                                                                                                                                                                                                                                                                                                                                                                                                                                                                                                                                                              \\ \hline

Object Detection Module Output  & {[}'camera', 'walking shoes', 'Ankle boot', 'Raincoat', 'light-colored dress', 'cotton clothes','water bottle', 'toothbrush', 'Binoculars'{]}

                \\ \hline

\end{tabular}
\end{table*}


\bibliographystyle{ACM-Reference-Format}
\bibliography{sample-base}

\end{document}